\documentclass[5p, longtitle]{elsarticle}

\usepackage{lineno,hyperref}

\usepackage{graphicx}
\usepackage{layout}
\usepackage{epstopdf}
\usepackage{amsmath}
\usepackage{amssymb}
\usepackage{verbatim}
\usepackage{multirow}
\usepackage{color}
\usepackage[utf8]{inputenc}
\usepackage{array}
\usepackage{makecell}
\usepackage{cellspace}
\usepackage{siunitx}
\usepackage{color}

\newcommand{\sign}{dd \to {}^4\text{He}\pi^0}

\newcommand{\bg}{dd \to {}^3\text{He}n\pi^0}
\newcommand{\bggg}{dd \to {}^3\text{He}n\gamma \gamma}
\newcommand{\direct}{dd \to {}^{4}\text{He} \gamma \gamma}
\newcommand{\directhyp}{dd \to {}^{4}\text{He} \gamma \gamma}
\newcommand{\mm}{dd \to {}^{4}\text{He}X}
\newcommand{\hyp}{dd \to {}^3\text{He}n\gamma \gamma}
\newcommand{\he}{{}^3\text{He}}
\newcommand{\al}{{}^4\text{He}}
\NewDocumentCommand{\varSI}{O{}}{\SI[math-rm=\mathnormal,parse-numbers=false,#1]}
\begin{document}

\begin{frontmatter}

\title{Importance of $d$-wave contributions in the charge symmetry breaking reaction $\sign$}

\author[IKPUU]{The WASA-at-COSY Collaboration\\[2ex] P.~Adlarson\fnref{fnmz}}
\author[ASWarsN]{W.~Augustyniak}
\author[IPJ]{W.~Bardan}
\author[Edinb]{M.~Bashkanov}
\author[MS]{F.S.~Bergmann}
\author[ASWarsH]{M.~Ber{\l}owski}
\author[Budker,Novosib]{A.~Bondar}
\author[PGI,DUS]{M.~B\"uscher}
\author[IKPUU]{H.~Cal\'{e}n}
\author[IFJ]{I.~Ciepa{\l}}
\author[PITue,Kepler]{H.~Clement}
\author[IPJ]{E.~Czerwi{\'n}ski}
\author[MS]{K.~Demmich}
\author[IKPJ]{R.~Engels}
\author[ZELJ]{A.~Erven}
\author[ZELJ]{W.~Erven}
\author[Erl]{W.~Eyrich}
\author[IKPJ,ITEP]{P.~Fedorets}
\author[Giess]{K.~F\"ohl}
\author[IKPUU]{K.~Fransson}
\author[IKPJ]{F.~Goldenbaum}
\author[IKPJ,IITI]{A.~Goswami}
\author[IKPJ,HepGat]{K.~Grigoryev}
\author[IKPUU]{C.--O.~Gullstr\"om}
\author[IKPJ,IASJ]{C.~Hanhart}
\author[IKPUU]{L.~Heijkenskj\"old\fnref{fnmz}}
\author[IKPJ]{V.~Hejny}
\author[MS]{N.~H\"usken}
\author[IPJ]{L.~Jarczyk}
\author[IKPUU]{T.~Johansson}
\author[IPJ]{B.~Kamys}
\author[ZELJ]{G.~Kemmerling\fnref{fnjcns}}
\author[IPJ]{G.~Khatri\fnref{fnharv}}
\author[MS]{A.~Khoukaz}
\author[IPJ]{O.~Khreptak}
\author[HeJINR]{D.A.~Kirillov}
\author[IPJ]{S.~Kistryn}
\author[ZELJ]{H.~Kleines\fnref{fnjcns}}
\author[Katow]{B.~K{\l}os}
\author[ASWarsH]{W.~Krzemie{\'n}}
\author[IFJ]{P.~Kulessa}
\author[IKPUU,ASWarsH]{A.~Kup{\'s}{\'c}}
\author[Budker,Novosib]{A.~Kuzmin}
\author[NITJ]{K.~Lalwani}
\author[IKPJ]{D.~Lersch}
\author[IKPJ]{B.~Lorentz}
\author[IPJ]{A.~Magiera}
\author[IKPJ,JARA]{R.~Maier}
\author[IKPUU]{P.~Marciniewski}
\author[ASWarsN]{B.~Maria{\'n}ski}
\author[ASWarsN]{H.--P.~Morsch}
\author[IPJ]{P.~Moskal}
\author[IKPJ]{H.~Ohm}
\author[IFJ]{W.~Parol}
\author[PITue,Kepler]{E.~Perez del Rio\fnref{fnlnf}}
\author[HeJINR]{N.M.~Piskunov}
\author[IKPJ]{D.~Prasuhn}
\author[IKPUU,ASWarsH]{D.~Pszczel}
\author[IFJ]{K.~Pysz}
\author[IKPUU,IPJ]{A.~Pyszniak}
\author[IKPJ,JARA,Bochum]{J.~Ritman}
\author[IITI]{A.~Roy}
\author[IPJ]{Z.~Rudy}
\author[IPJ]{O.~Rundel}
\author[IITB]{S.~Sawant}
\author[IKPJ]{S.~Schadmand}
\author[IPJ]{I.~Sch\"atti--Ozerianska}
\author[IKPJ]{T.~Sefzick}
\author[IKPJ]{V.~Serdyuk}
\author[Budker,Novosib]{B.~Shwartz}
\author[MS]{K.~Sitterberg}
\author[PITue,Kepler,Tomsk]{T.~Skorodko}
\author[IPJ]{M.~Skurzok}
\author[IPJ]{J.~Smyrski}
\author[ITEP]{V.~Sopov}
\author[IKPJ]{R.~Stassen}
\author[ASWarsH]{J.~Stepaniak}
\author[Katow]{E.~Stephan}
\author[IKPJ]{G.~Sterzenbach}
\author[IKPJ]{H.~Stockhorst}
\author[IKPJ,JARA]{H.~Str\"oher}
\author[IFJ]{A.~Szczurek}
\author[ASWarsN]{A.~Trzci{\'n}ski}
\author[IKPUU]{M.~Wolke}
\author[IPJ]{A.~Wro{\'n}ska}
\author[ZELJ]{P.~W\"ustner}
\author[KEK]{A.~Yamamoto}
\author[ASLodz]{J.~Zabierowski}
\author[IPJ]{M.J.~Zieli{\'n}ski}
\author[IKPUU]{J.~Z{\l}oma{\'n}czuk}
\author[ASWarsN]{P.~{\.Z}upra{\'n}ski}
\author[IKPJ]{M.~{\.Z}urek\corref{coau}}\ead{m.zurek@fz-juelich.de}

\address[IKPUU]{Division of Nuclear Physics, Department of Physics and 
 Astronomy, Uppsala University, Box 516, 75120 Uppsala, Sweden}
\address[ASWarsN]{Department of Nuclear Physics, National Centre for Nuclear 
 Research, ul.\ Hoza~69, 00-681, Warsaw, Poland}
\address[IPJ]{Institute of Physics, Jagiellonian University, prof.\ 
 Stanis{\l}awa {\L}ojasiewicza~11, 30-348 Krak\'{o}w, Poland}
\address[Edinb]{School of Physics and Astronomy, University of Edinburgh, 
 James Clerk Maxwell Building, Peter Guthrie Tait Road, Edinburgh EH9 3FD, 
 Great Britain}
\address[MS]{Institut f\"ur Kernphysik, Westf\"alische Wilhelms--Universit\"at 
 M\"unster, Wilhelm--Klemm--Str.~9, 48149 M\"unster, Germany}
\address[ASWarsH]{High Energy Physics Department, National Centre for Nuclear 
 Research, ul.\ Hoza~69, 00-681, Warsaw, Poland}
\address[Budker]{Budker Institute of Nuclear Physics of SB RAS, 11~akademika 
 Lavrentieva prospect, Novosibirsk, 630090, Russia}
\address[Novosib]{Novosibirsk State University, 2~Pirogova Str., Novosibirsk, 
 630090, Russia}
\address[PGI]{Peter Gr\"unberg Institut, PGI--6 Elektronische Eigenschaften, 
 Forschungszentrum J\"ulich, 52425 J\"ulich, Germany}
\address[DUS]{Institut f\"ur Laser-- und Plasmaphysik, Heinrich--Heine 
 Universit\"at D\"usseldorf, Universit\"atsstr.~1, 40225 D\"usseldorf, Germany}
\address[IFJ]{The Henryk Niewodnicza{\'n}ski Institute of Nuclear Physics, 
 Polish Academy of Sciences, Radzikowskiego~152, 31--342 Krak\'{o}w, Poland}
\address[PITue]{Physikalisches Institut, Eberhard--Karls--Universit\"at 
 T\"ubingen, Auf der Morgenstelle~14, 72076 T\"ubingen, Germany}
\address[Kepler]{Kepler Center f\"ur Astro-- und Teilchenphysik, 
 Physikalisches Institut der Universit\"at T\"ubingen, Auf der 
 Morgenstelle~14, 72076 T\"ubingen, Germany}
\address[IKPJ]{Institut f\"ur Kernphysik, Forschungszentrum J\"ulich, 52425 
 J\"ulich, Germany}
\address[ZELJ]{Zentralinstitut f\"ur Engineering, Elektronik und Analytik, 
 Forschungszentrum J\"ulich, 52425 J\"ulich, Germany}
\address[Erl]{Physikalisches Institut, Friedrich--Alexander--Universit\"at 
 Erlangen--N\"urnberg, Erwin--Rommel-Str.~1, 91058 Erlangen, Germany}
\address[ITEP]{Institute for Theoretical and Experimental Physics named 
 by A.I.\ Alikhanov of National Research Centre ``Kurchatov Institute'', 
 25~Bolshaya Cheremushkinskaya, Moscow, 117218, Russia}
\address[Giess]{II.\ Physikalisches Institut, Justus--Liebig--Universit\"at 
 Gie{\ss}en, Heinrich--Buff--Ring~16, 35392 Giessen, Germany}
\address[IITI]{Department of Physics, Indian Institute of Technology Indore, 
 Khandwa Road, Simrol, Indore - 453552, Madhya Pradesh, India}
\address[HepGat]{High Energy Physics Division, Petersburg Nuclear Physics 
 Institute named by B.P.\ Konstantinov of National Research Centre ``Kurchatov 
 Institute'', 1~mkr.\ Orlova roshcha, Leningradskaya Oblast, Gatchina, 188300, 
 Russia}
\address[IASJ]{Institute for Advanced Simulation, Forschungszentrum J\"ulich, 
 52425 J\"ulich, Germany}
\address[HeJINR]{Veksler and Baldin Laboratory of High Energiy Physics, 
 Joint Institute for Nuclear Physics, 6~Joliot--Curie, Dubna, 141980, Russia}
\address[Katow]{August Che{\l}kowski Institute of Physics, University of 
 Silesia, Uniwersytecka~4, 40--007, Katowice, Poland}
\address[NITJ]{Department of Physics, Malaviya National Institute of 
 Technology Jaipur, JLN Marg Jaipur - 302017, Rajasthan, India}
\address[JARA]{JARA--FAME, J\"ulich Aachen Research Alliance, Forschungszentrum
 J\"ulich, 52425 J\"ulich, and RWTH Aachen, 52056 Aachen, Germany}
\address[Bochum]{Institut f\"ur Experimentalphysik I, Ruhr--Universit\"at 
 Bochum, Universit\"atsstr.~150, 44780 Bochum, Germany}
\address[IITB]{Department of Physics, Indian Institute of Technology Bombay, 
 Powai, Mumbai - 400076, Maharashtra, India}
\address[Tomsk]{Department of Physics, Tomsk State University, 36~Lenina 
 Avenue, Tomsk, 634050, Russia}
\address[KEK]{High Energy Accelerator Research Organisation KEK, Tsukuba, 
 Ibaraki 305--0801, Japan}
\address[ASLodz]{Astrophysics Division, National Centre for Nuclear Research, 
 Box~447, 90--950 {\L}\'{o}d\'{z}, Poland}

\fntext[fnmz]{present address: Institut f\"ur Kernphysik, Johannes 
 Gutenberg--Universit\"at Mainz, Johann--Joachim--Becher Weg~45, 55128 Mainz, 
 Germany}
\fntext[fnjcns]{present address: J\"ulich Centre for Neutron Science JCNS, 
 Forschungszentrum J\"ulich, 52425 J\"ulich, Germany}
\fntext[fnharv]{present address: Department of Physics, Harvard University, 
 17~Oxford St., Cambridge, MA~02138, USA}
\fntext[fnlnf]{present address: INFN, Laboratori Nazionali di Frascati, Via 
 E.~Fermi, 40, 00044 Frascati (Roma), Italy}

\cortext[coau]{Corresponding author}

\begin{abstract}
This letter reports a first quantitative analysis of the contribution of higher partial waves in the charge symmetry breaking reaction $\sign$ using the WASA-at-COSY detector setup at an excess energy of $Q = \SI{60}{MeV}$. The determined differential cross section can be parametrized as $\text{d}\sigma/\text{d}\Omega = a + b\cos^{2}\theta^*$, where $\theta^*$ is the production angle of the pion in the center-of-mass coordinate system, and the results for the parameters are $a = \left(1.55 \pm  0.46 (\text{stat}) ^{+0.32}_{-0.8} (\text{syst}) \right)$\,\si{pb\per sr} and $b = \left(13.1 \pm 2.1 (\text{stat}) ^{+1.0}_{-2.7} (\text{syst})\right)$\,\si{pb\per sr}. The data are compatible with vanishing $p$-waves and a sizable $d$-wave contribution. This finding should strongly constrain the contribution of the $\Delta$ isobar to the $\sign$ reaction and is, therefore, crucial for a quantitative understanding of quark mass effects in nuclear production reactions.
\end{abstract}


\end{frontmatter}


\section{Introduction}

Within the Standard Model of elementary particles isospin symmetry is violated via quark mass differences as well as electromagnetic effects~\cite{Weinberg:1977hb,Gasser:1982ap,Leutwyler:1996qg}. On the hadronic level this is reflected, for example, by the proton-neutron mass difference. It is due to quark-mass effects that the proton is lighter than the neutron and, therefore, stable. The observation of isospin violation (IV) in hadronic reactions in principle allows one to study the effects of quark masses. However, most experimental signatures of IV are dominated by the pion mass difference $m_{\pi^0} - m_{\pi^{\pm}}$, which is to a very good approximation of purely electromagnetic origin. An exception are observables that are charge symmetry breaking (CSB). Charge symmetry, a subgroup of isospin symmetry, is the invariance of the Hamiltonian under rotation by 180$^\circ$ around the second axis in isospin space that interchanges up and down quarks. The charge symmetry operator does not interchange charged and neutral pion states, and the pion mass difference does not enter (see, e.g.,~\cite{Miller:1990iz}). On the basis of theoretical approaches with a direct connection to QCD, like lattice QCD and chiral perturbation theory (ChPT), it is, therefore, possible to link quark-mass effects to hadronic observables. 

While CSB observables have the advantage of being directly related to quark-mass differences, their smallness poses an experimental challenge. First precision measurements of CSB were reported for the reaction $\sign$ at beam energies very close to the reaction threshold~\cite{Stephenson:2003dv} and, at the same time, via a non-vanishing forward-back\-ward asymmetry in $np \to d \pi^{0}$~\cite{Opper:2003sb}. Both results triggered a series of theoretical investigations. The signal of the latter measurement was shown to be proportional to the quark-mass-induced part of the proton-neutron mass difference up to next-to-leading order in ChPT~\cite{Miller:2006tv,Filin:2009yh}. This became possible by the adaption of ChPT to pion production reactions in Ref.~\cite{Hanhart2004155}. The formalism has recently been pushed to next-to-next-to-leading order for $s$-waves~\cite{Filin:2013uma,Baru:2016kru}. The contribution of $p$-waves has been investigated in Ref.~\cite{Baru:2009fm}. For a recent review see Ref.~\cite{Baru:2013zpa}. 

For the reaction $\sign$ the four-nucleon interaction in initial and final state adds an additional facet. First steps towards a theoretical understanding of this reaction were taken in Refs.~\cite{Gardestig:2004hs,Nogga:2006cp}. Additional CSB effects from soft photons in the initial state have been studied in Refs.~\cite{Lahde:2007nu,Fonseca:2009qs}. The focus in that work has been on $s$-waves in the final state, since no experimental information on higher partial waves was available at that time. However, such information is important, since it will allow one to constrain the contribution from the $\Delta$ resonance that is known to provide the bulk of the $p$-wave contributions in the isospin conserving $pp \to d\pi^+$ reaction~\cite{Niskanen:1978vm,Niskanen:1995wy,Hanhart:1998za} --- without this, a quantitative control of higher order operators for the reaction at hand appears impossible.
A first measurement with WASA was inconclusive due to limited statistics \cite{4He}. Thus, there are no theoretical predictions for higher energies and/or higher partial waves yet.
In this paper, data are presented for the first time that quantify the contribution of higher partial waves to the reaction $\sign$. 

\section{Experiment}

The ten-week-long experiment was performed at the Cooler Synchrotron COSY \cite{Maier:1997} of the Institute for Nuclear Physics at the Forschungszentrum J\"ulich in Germany. The particles produced in the collisions of a deuteron beam with a momentum of $p_d = \SI{1.2}{GeV\per \textit{c}}$ ($Q = \SI{60}{MeV}$) with frozen deuteron pellets were detected in the modified WASA facility \cite{Proposal_WASA}. The setup consisted of forward and central detectors, where the $\al$ ejectiles and the photons from the $\pi^0$ decay were detected, respectively. For this experiment the forward detector was optimized for a time-of-flight (TOF) measurement. Several layers of the original detector were removed to introduce a free flight path of more than $\SI{1.5}{m}$. This modification provides access to an additional, independent observable for energy calibration and particle selection --- in the previous measurement~\cite{4He} these were based only on the correlation of energy losses in the detector layers. The new setup consisted of an array of straw tubes for precise tracking and three layers of plastic scintillators for energy reconstruction and particle identification: two $\SI{3}{mm}$ thick layers of the forward window counter, used as start detectors, and the $\SI{20}{mm}$ thick layer of the forward veto hodoscope, used as a stop detector. Photons from the $\pi^0$ decay were detected in the central electromagnetic calorimeter and discriminated from charged particles by means of a veto signal from the plastic scintillator barrel located inside the calorimeter.

The main trigger required a high energy deposit in at least one element of the first and the second layer of the forward window counter and at least one cluster originating from a neutral particle in the central detector.

\section{Analysis}

The signature of the $\sign$ reaction is a forward-going $\al$ particle and two photons from the decay of the $\pi^0$. The only other channel with $\al$ and two photons in the final state is the double radiative capture reaction $\direct$ as an irreducible physics background. A further source of background is the isospin symmetry conserving $\bg$ reaction with a more than four orders of magnitude larger cross section \cite{3He}. The suppression of this reaction is challenging since $\he$ and $\al$ have similar, given the detector resolution, energy losses in the forward window counters. Compared to $\bg$, the direct two photon production in $\bggg$ is suppressed by a factor of $\alpha^2$ (with $\alpha$ being the fine-structure constant) and can be neglected.

The energy loss in the forward window counters and TOF have been used to reconstruct the kinetic energy of the outgoing $\he$ and $\al$ particles by matching their patterns to Monte Carlo simulations. The full four-vectors have been obtained using in addition the azimuthal and polar angles reconstructed by the forward tracking detector. For the further analysis at least one track in the forward detector and at least two reconstructed clusters of crystals with energy deposited by neutral particles in the central detector have been required.

The final candidate events have been selected by means of a kinematic fit. The purpose of the fit was to improve the precision of the measured kinematic variables and to serve as a selection criterion for background reduction. For the assumed reaction hypothesis the measured variables were varied within the experimental uncertainties until certain kinematic constraints were fulfilled, here the overall momentum and energy conservation. For every event the $\hyp$ and $\directhyp$ hypotheses have been tested separately. No additional constraint on the invariant mass of the two photons has been imposed, in order to be able to measure the signal yield using the two-photon invariant-mass distribution. In case of more than one track in the forward detector or more than two neutral clusters in the central detector (caused by event pileup or low energy satellites of the main photon clusters) the combination with the smallest $\chi^2$ from the fit has been chosen.

The reduction of the $\bg$ background by four orders of magnitude has been mainly achieved using a cut on the two-dimensional cumulative probability distribution from the kinematic fits, analogously as described in Ref.~\cite{4He}. The cut has been optimized by maximizing the statistical significance of the $\pi^0$ signal in the final missing mass plot. 

\begin{figure}
\includegraphics[width=86mm]{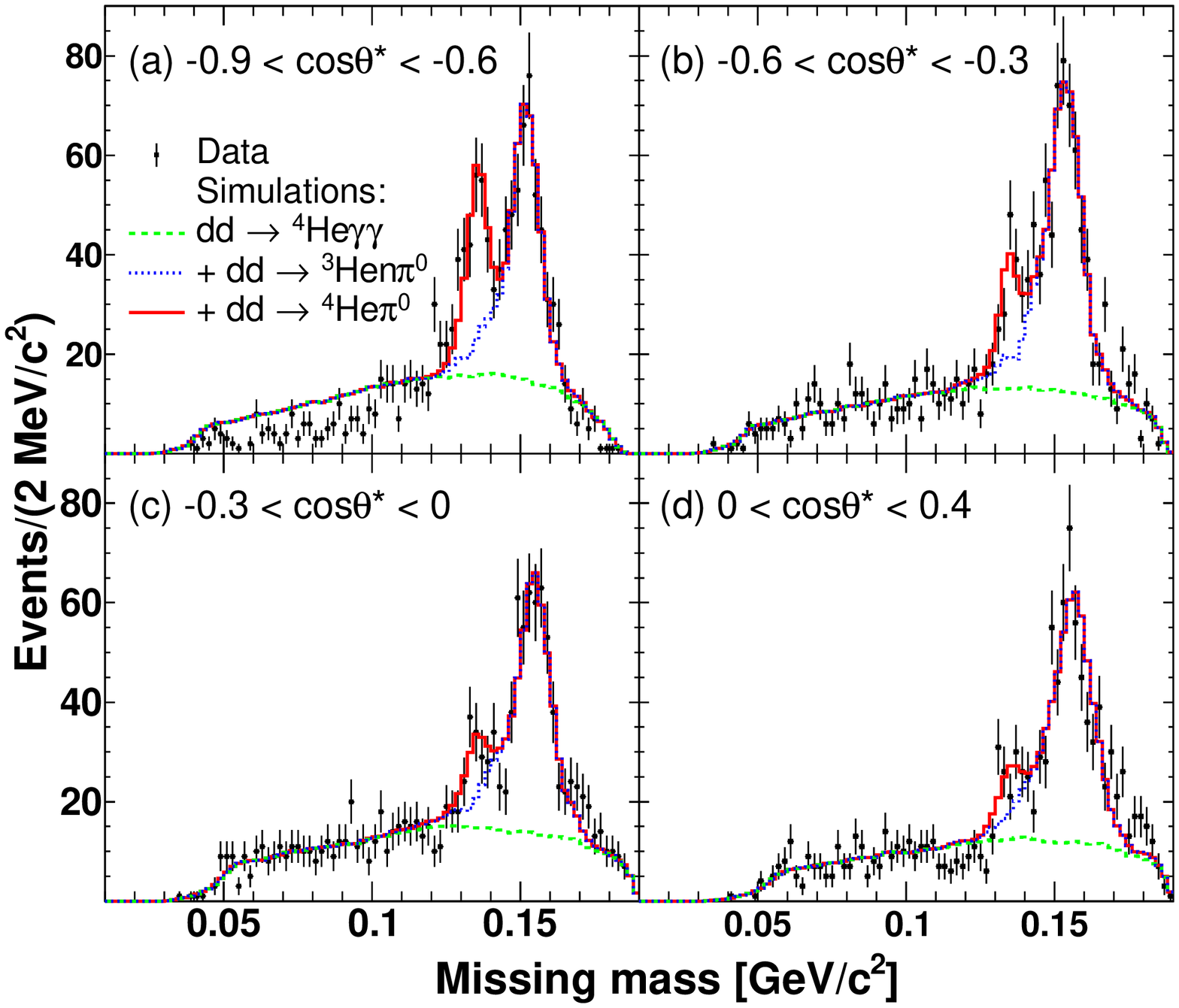}
\caption{Missing mass for the $\mm$ reaction for the four angular bins of the production angle of the pion in the center-of-mass system. The spectrum is fitted with a linear combination of the simulated signal and background reactions: double radiative capture $\direct$ (green dashed line), plus $\bg$ (blue dotted line), plus $\sign$ (red solid line). The fit excludes the missing mass region below $\SI{0.11}{GeV\per \textit{c}^2}$.}\label{fig:MissMass}
\end{figure}

The four-momenta obtained from the kinematic fit of the $\directhyp$ hypothesis have been used to calculate the missing mass $m_{X}$ for the reaction $\mm$ as a function of the center-of-mass production angle $\theta^*$ of the $\pi^0$. In Fig.~\ref{fig:MissMass} the missing mass spectra for the four angular bins within the detector acceptance ($-0.9 \leq \cos\theta^{*} \leq 0.4$) are presented. On a smooth background from double radiative capture $\direct$ two significant peaks are visible. One of these, originating from the signal reaction $\sign$, is located at the $\pi^0$ mass of $\SI{0.135}{GeV\per \textit{c}^2}$. The other one corresponds to misidentified events from the background reaction $\bg$ and is shifted by the ${}^3\text{He} - n$ binding energy. The missing mass spectra have been fitted with a linear combination of the following high-statistics Monte Carlo templates: (i) $\direct$ assuming a 3-body phase-space distribution, (ii) $\bg$ using the model from \cite{3He}, and (iii) the two-body reaction $\sign$. For each $\cos\theta^*$ bin, a fit of the Monte Carlo templates to the data has been performed with the constraint that the sum of the fitted templates has to fit the overall missing mass spectrum. As result, the $\pi^0$ peak from the $\sign$ reaction contains $336 \pm 43$ events in total.

 In the course of the fit the Monte Carlo templates have been modified in two ways. In the missing mass spectra, the background originating from misidentified $\bg$ is slightly shifted in comparison to data. This shift can be attributed to systematic differences in the simulated detector response for $\al$ and misidentified $\he$. With a cut efficiency close to $10^{-4}$ the latter mainly originate from the tails of the corresponding distributions. Nevertheless, the shape of background contribution is well described. Therefore, this mismatch has been compensated by introducing an angle-dependent scaling factor in the missing mass $m_X$ as free parameter. The obtained factors (from backward to forward angles) are within the range of 1.005--0.972. The second modification concerns the missing mass spectrum below $\SI{0.11}{GeV\per \textit{c}^2}$ in the most backward angular bin. This region is dominated by the $\direct$ reaction, which has been simulated using 3-body phase space. This model, however, underestimates the contribution in that region. The dominating background from the $\bg$ reaction at higher missing masses prevents describing all contributions precisely enough to verify more advanced models. For a consistent description in all angular bins, for the final fit the missing mass range below $\SI{0.11}{GeV\per \textit{c}^2}$ has been excluded in all angular bins.

For the final acceptance correction, the $\sign$ generator with the angular distribution obtained in this analysis has been used. The integrated luminosity has been calculated using the $\bg$ reaction, based on a measurement with WASA at $p_d = \SI{1.2}{GeV\per \textit{c}}$ \cite{3He}. It equals to $(37.2 \pm 3.7 (\text{norm}) \pm 0.1 (\text{syst}))$\,\si{pb^{-1}}, which is about 7.5 times larger than the value from the previous measurement with WASA reported in Ref.~\cite{4He}

The stability of the results has been tested against variations of all selection cuts, according to method described in Ref.~\cite{Barlow:2002}. Out of these, the only statistically significant effect has been observed with the variation of the cumulative probability distribution cut and added as systematic uncertainty. The sensitivity of the overall fit has been checked by varying the fit parameters, especially the linear scaling factor in $m_X$, and using smooth analytic functions to reproduce the shape of background at low missing masses. No significant change in the result has been observed while maintaining the goodness-of-fit in the peak region. Thus, no systematic uncertainty has been assigned here. The error on the normalization to the $\bg$ reaction has been propagated to the final result.

\section{Results}
\begin{figure}
\includegraphics[width=86mm]{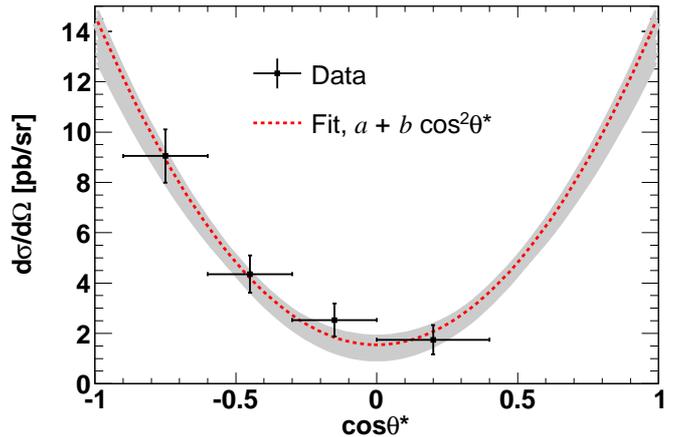}
\caption{Angular distribution of the $\sign$ reaction at $Q = \SI{60}{MeV}$. The result of the fit up to second order in $\cos\theta^*$ is shown with a dotted curve. The systematic errors of the fit are presented as a gray band. The horizontal error bars indicate the bin width.}
\label{fig:AngDist}
\end{figure}

Figure~\ref{fig:AngDist} presents the obtained differential cross section. Since identical particles in the initial state require a forward-backward symmetric cross section, it has been fitted using the function $\text{d}\sigma/\text{d}\Omega = a + b\cos^{2}\theta^*$ resulting in:
\begin{subequations}
\begin{align} 
 &a = \left(1.55 \pm  0.46 (\text{stat}) ^{+0.32}_{-0.8} (\text{syst})\right) \si{pb/sr},\\
 &b = \left(13.1  \pm 2.1 (\text{stat}) ^{+1.0}_{-2.7} (\text{syst})\right) \si{pb/sr}.
 \label{Eq:bfit}
\end{align}
\end{subequations}
Both parameters have an additional, common systematic uncertainty of about $10\%$ from normalization. 

\begin{figure}
\includegraphics[width=86mm]{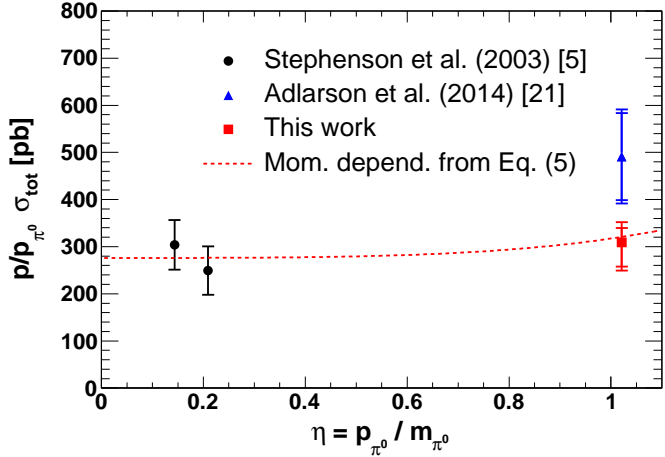}
\caption{The $\sign$ reaction amplitude squared $(p/p_{\pi^0})\sigma_{\text{tot}}$ as a function of $\eta = p_{\pi^0}/m_{\pi^0}$. The circles represent the results from \cite{Stephenson:2003dv}, the square corresponds to the final result for the total cross section from this work, and the triangle represents the cross section from the previous WASA measurement \cite{4He}. Note that the result from \cite{4He} has been obtained assuming pure $s$-wave. The error bars show the combined statistical and systematic uncertainties. For the results obtained with WASA the error bars with subtracted common uncertainty originating from normalization are also presented. The dotted curve indicates the momentum dependence of the total cross section from Eq.~(\ref{Eq:s}) with the fitted amplitudes from Eq.~(\ref{Eq:A2C}).}
\label{fig:CSmom}
\end{figure}

The total cross section obtained as the integral of the function fitted to the angular distribution amounts to:
\begin{equation}
\sigma_{\text{tot}} = (74.3 \pm 6.8 (\text{stat}) ^{+1.2}_{-10.1}(\text{syst}) \pm 7.7 (\text{norm})) \text{pb}.
\label{Eq:totcs}
\end{equation}
Fig.~\ref{fig:CSmom} shows the resulting momentum dependence of the reaction amplitude $(p/p_{\pi^0})\sigma_{\text{tot}}$ including the data from Ref.~\cite{Stephenson:2003dv}.
Here, $p_{\pi^0}$ is the momentum of the pion and $p$ is the incident-deuteron momentum, both in the center-of-mass system.

The cross sections are systematically smaller than the results reported in Ref.~\cite{4He}, however, consistent within errors. In view of the limited statistics a decisive analysis of this difference is difficult. As most probable reason our studies identified the implementation of nuclear interactions of $\he$ in the Monte Carlo simulations. It was found that this effect was not properly taken into account in the analysis of the previous data. This resulted in an increased (simulated) detection efficiency for the normalization reaction and, consequently, in a too low luminosity. As the effect is the largest for the stopping layer, the analysis of the current data set is less sensitive as it is based on a TOF measurement and does not rely on energy correlations only.

For a further analysis of the differential cross section in terms of partial waves in the final state, the formalism from Ref.~\cite{Wronska:2005wk} has been used. Considering only $s$- and $p$-waves the parameter $b$ can be written as:
\begin{equation}
b = - \frac{p_{\pi^0}}{p}\frac{2}{3}|C|^2p_{\pi^0}^2, 
\label{Eq:b}
\end{equation}
where $C$ is the $p$-wave amplitude. Note that the symmetry of the initial state requires that only partial waves of the same parity interfere. Up to this order, $p$-waves contribute with a negative sign corresponding to a maximum at $\theta^{*} = 90^{\circ}$ in the angular distribution. The observed minimum can only be explained extending the formalism to $d$-waves in the final state. Therefore, these data establish for the first time the presence of a sizable contribution of $d$-waves to the $\sign$ reaction, which have so far not been considered in the theoretical calculations.

A consistent description that includes $d$-waves has to consider terms up to fourth order in pion momentum. Following Ref.~\cite{Wronska:2005wk} the differential cross section can be written as:
\begin{equation}
\begin{aligned}
\frac{\text{d}\sigma}{\text{d}\Omega} = {} &\frac{p_{\pi^0}}{p}\frac{2}{3}\Bigl( |A_0|^2 + 2\operatorname{Re}(A_0^*A_2)P_2(\cos\theta^*)p_{\pi^0}^2 \Bigr. \\ 
 & + |A_2|^2P_2^2(\cos\theta^*)p_{\pi^0}^4 + |C|^2\sin^2\theta^*p_{\pi^0}^2 \\
 &  + \Bigl.|B|^2\sin^2\theta^*\cos^2\theta^* p_{\pi^0}^4 \Bigr).\label{Eq:dsdO}
\end{aligned}
\end{equation}
Here, $A_0$ is the $s$-wave amplitude, $A_2$ and $B$ are the $d$-wave amplitudes, and $P_2$ is the second order Legendre polynomial. The corresponding expression for the total cross section reads:
\begin{equation}
\begin{aligned}
\sigma_{\text{tot}} = {} &\frac{p_{\pi^0}}{p}\frac{8\pi}{3}\Bigl(|A_0|^2 + \frac{2}{3}|C|^2 p_{\pi^0}^2 \Bigr.\\
& + \frac{1}{5}|A_2|^2 p_{\pi^0}^4 + \Bigl.\frac{2}{15}|B|^2 p_{\pi^0}^4 \Bigr).
 \label{Eq:s}
 \end{aligned}
\end{equation}
Since a full fit with four independent amplitudes and one relative phase is outside the scope of the presented data, quantitative results can only be obtained using additional constraints. An unbiased determination of the amplitudes is not possible under these circumstances, thus, the focus is on the correlations between them.

If one assumes that the amplitude $A_0$ does not carry any momentum dependence, it can be extracted from the results in Ref.~\cite{Stephenson:2003dv} where $s$-wave is by far dominating. The obtained value is $|A_0|_{\text{thr}} = \left(5.74 \pm 0.38(\text{stat})\right)$\,$\left(\text{pb/sr}\right)^{1/2}$, which can then be used as fixed parameter in the fit of the angular distribution at $Q = \SI{60}{MeV}$. Furthermore, systematic studies of the behaviour of the fit with respect to $B$ and the relative phase $\delta$ between $A_0$ and $A_2$ (i.e., $\Re\{A_0^*A_2\} = |A_0||A_2|\cos\delta$) show that the data are not sensitive to $|B|$ and $\delta$, which have comparatively large errors and are consistent with zero. For example, the fit with the parameters $|A_2|$, $|B|$, $|C|$ free and $\delta$ fixed to zero results in $|B| = \left(150 ^{+130}_{-420}(\text{stat})\right)$ $\left(\text{pb/sr}\right)^{1/2}(\text{GeV}/c)^{-2}$, and the fit with $|A_2|$, $\delta$, $|C|$ free and $|B|$ fixed to zero results in $\delta = 0 \pm 0.66(\text{stat})$. Moreover, the parameters $|C|$ and $|A_2|$ from both fits are consistent within the uncertainties. Consequently, both $|B|$ and $\delta$ have been fixed to zero.

From the final fit of the angular distribution at $Q = \SI{60}{MeV}$ with all described constraints the following amplitudes have been extracted:

\begin{subequations}
\begin{align}
|A_{2}| =& \left( 258 ^{+50}_{-42}(\text{stat}) ^{+45}_{-38} (\text{syst}) \right.\nonumber \\ &\left.^{+37}_{-12} (\text{norm}) \right) \frac{\left(\text{pb/sr}\right)^{1/2}}{(\text{GeV}/c)^{2}},\label{Eq:A2} \allowdisplaybreaks\\
|C| =& \left( 6 ^{+9}_{-21} (\text{stat}) ^{+3}_{-10} (\text{syst}) \right.\nonumber \\ &\left.^{+10}_{-5} (\text{norm}) \right)  \frac{ \left(\text{pb/sr}\right)^{1/2}}{\text{GeV}/c}.\label{Eq:C}
\end{align}
\label{Eq:A2C}
\end{subequations}
The asymmetric statistical errors are a consequence of the non-linearity of the fit function.  

\begin{figure}
\includegraphics[width=86mm]{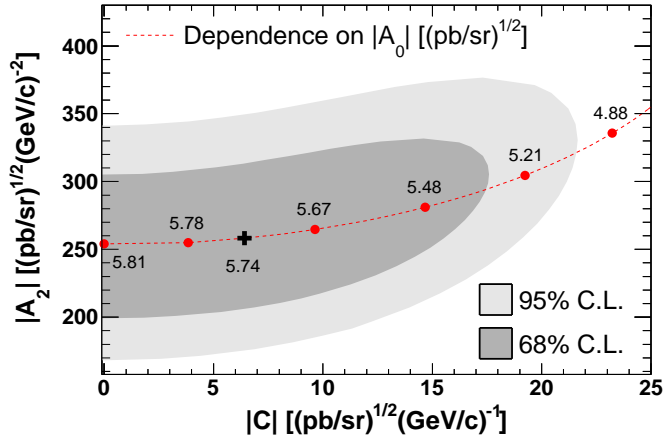}
\caption{Correlation plot for the parameters $|C|$ and $|A_2|$ determined from the fit of the angular distribution of $\sign$ at $Q = \SI{60}{MeV}$. The center point marked with the cross shows the result from Eq.~(\ref{Eq:A2C}). The shaded areas indicate the 68\% and 95\% confidence regions. The dotted line shows the influence of a variation of $|A_{0}|$ on $|C|$ and $|A_2|$, with the circle points representing the results for the indicated values of $|A_{0}|$.}
\label{fig:CL}
\end{figure}

\begin{figure}
\includegraphics[width=86mm]{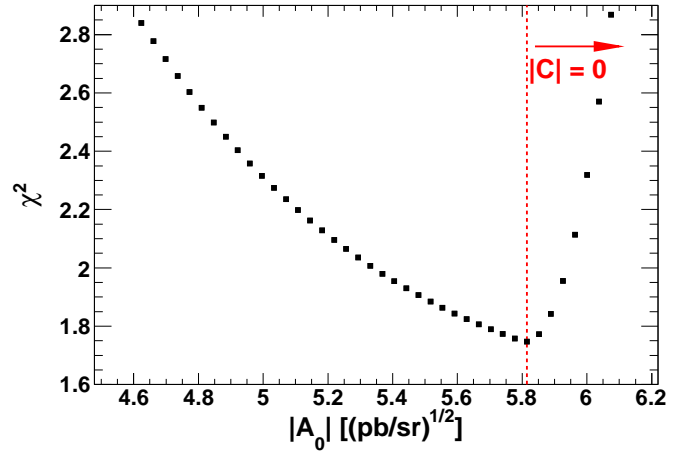}
\caption{Minimal total $\chi^2$ from the fit of the angular distribution of $\sign$ at $Q = \SI{60}{MeV}$ as a function of the fixed value of the $s$-wave amplitude $|A_0|$. The dotted line indicates the value of $|A_0|$ for which the $p$-wave contribution given by the parameter $|C|$ vanishes. A further increase of $|A_0|$ still keeps $|C|$ at 0 at the cost of the goodness-of-fit.}
\label{fig:chi2}
\end{figure}

Figure~\ref{fig:CL} shows a correlation plot between the parameters $|C|$ and $|A_2|$. The center point marked with a cross shows the result from Eq.~(\ref{Eq:A2C}). The shaded areas indicate the 68\% and 95\% confidence regions. The dotted line shows the dependence of the central values for $|C|$ and $|A_2|$ on $|A_0|$ --- some values for $|A_0|$ are shown explicitly in the figure. The minimal total $\chi^2$ as a function of the fixed value of $|A_0|$ is presented in Fig.~\ref{fig:chi2}. At $|A_0| =  5.81\,\left(\text{pb/sr}\right)^{1/2}$ the $p$-wave contribution given by the parameter $|C|$ vanishes. A further increase of $|A_0|$ still keeps $|C|$ at 0 at the cost of the goodness-of-fit. One can see that the fit to the data has the tendency to maximize $|A_0|$ and, thus, minimize $|C|$. This maximum value of $|A_0|$ is consistent with the one obtained from Ref.~\cite{Stephenson:2003dv} supporting the assumption of a momentum independent $s$-wave amplitude. Furthermore, when $|C|$ vanishes and $|A_{0}|$ has its maximum value, the corresponding minimal $|A_2|$ value still significantly differs from zero. Even if one allows $|A_0|$ to drop with increasing momentum, this is compensated by larger values of $|C|$ to maintain the total cross section. At the same time the value of $|A_2|$ also increases, i.e., the $d$-wave contribution would become even larger.

\section{Summary}

In summary, this letter reports for the first time a successful measurement of higher partial waves in the differential cross section of the charge symmetry violating reaction $\sign$. The data with a minimum at $\theta^{*} = 90^{\circ}$ can be understood only by the presence of a significant $d$-wave contribution in the final state. At the same time they are consistent with a vanishing $p$-wave. Existing theoretical calculations to describe the reaction $\sign$ within Chiral Perturbation Theory are limited to $s$-wave pion production. There are first considerations to extend these efforts to $p$-waves in the final state, however, the presented data show that this is not sufficient.  

It is well known from phenomenology as well as studies using effective field theory that the $\Delta$ isobar plays a crucial role in pion production reactions, especially for partial waves higher than $s$-wave~\cite{Niskanen:1978vm,Niskanen:1995wy,Hanhart:1998za}. Since isospin conservation does not allow for the excitation of a single $\Delta$ in the $dd$ state, the appearance of prominent higher partial waves in $\sign$ might point at an isospin violating excitation of the $\Delta$ isobar. This indicates that a theoretical analysis of the data presented in the letter should allow for deep insights not only into the dynamics of the nucleon-nucleon interaction but also into the role of quark masses in hadron dynamics.

\section*{Acknowledgements}

We would like to thank the technical staff of the COoler SYnchrotron COSY. We thank C.~Wilkin for valuable discussions. This work was supported in part by the EU Integrated Infrastructure Initiative Hadron Physics Project under contract number RII3-CT-2004-506078; by the European Commission under the 7th Framework Programme through the Research Infrastructures action of the Capacities Programme, Call: FP7-INFRASTRUCTURES-2008-1, Grant Agreement No. 227431; by the Polish National Science Centre through the grants 2016/23/B/ST2/00784, 2014/15/N/ST2/03179, DEC-2013/11/N/ST2/04152, and the Foundation for Polish Science (MPD), co–financed by the European Union within the European Regional Development Fund. We acknowledge the support given by the Swedish Research Council, the Knut and Alice Wallenberg Foundation, and the Forschungszentrum J\"ulich FFE Funding Program. This work is based on the PhD thesis of Maria \.Zurek.

\section*{References}
\bibliography{mybibfile}

\end{document}